\documentclass[aps,prl,twocolumn,showkeys,amsmath,amssymb,superscriptaddress,nofootinbib,floatfix,longbibliography]{revtex4-2}
\usepackage{graphicx,color,ulem}
\usepackage{epsfig,bm,slashed,hyperref}

\begin{document}

\title{Wounded parton scaling of multiplicities in ultra-relativistic \\ light- and heavy-ion collisions
}

\author{Rupam Samanta}
\email{rupam.samanta@ifj.edu.pl}
\affiliation{Institute of Nuclear Physics, Polish Academy of Sciences,  31-342 Krak\'ow, Poland}

\author{Piotr Bo\.zek}
\email{piotr.bozek@fis.agh.edu.pl}
\affiliation{AGH University of Krak\'ow, Faculty of Physics and Applied Computer Science, 30-059 Krak\'ow, Poland}

\author{Wojciech Broniowski}
\email{wojciech.broniowski@ifj.edu.pl}
\affiliation{Institute of Nuclear Physics, Polish Academy of Sciences,  31-342 Krak\'ow, Poland}

\begin{abstract}
Multiplicities of charged particles produced in O+O, Ne+Ne, Xe+Xe, and Pb+Pb collisions at $\sqrt{s_{NN}} \sim 5$~TeV are studied in a uniform way within a wounded parton Glauber framework with 
overlaid negative binomial fluctuations. In this model, the nucleon's inelastic interaction is modeled via its constituent partons, 
whose number is a parameter, with best description obtained with four partons per nucleon.   
We fit directly the experimental multiplicity distributions (histograms), using {\it the same model parameters} for each reaction. The fit is performed in the 
c=1–80\% centrality range. Avoiding the most peripheral events makes the method insensitive to the normalization issues caused by the difficulty in separating the Coulomb interactions, 
whereas the most central collisions may involve a different particle production mechanism. We find a proper
model description of multiplicity distributions across all the studied systems for $c \lesssim 1\%$.
\end{abstract}


\maketitle

{\it \bf Introduction.} Multiplicity of produced particles is by far the simplest measured quantity in ultra-relativistic nuclear collisions; it reflects the total 
entropy deposition in the initially formed fireball, which at later stages undergoes expansion via a viscous hydrodynamic evolution of the 
thermalized quark-gluon plasma, finally freezing out into hadrons~\cite{Ollitrault:2010tn,Heinz:2013th,Busza:2018rrf,Florkowski:2010zz}. 
Recently, the so called light-ion collisions (with $^{16}$O and $^{20}$Ne beams) were investigated at the CERN Large Hadron Collider (LHC)~\cite{ATLAS:2025nnt,ATLAS:2026zgq,ALICE:2025luc,ALICE:2026zck,CMS:2025tga,CMS:2025bta,CMS:2026qef,CMS:2026sai}.

In this Letter we show that a {\it universal}  description of the bulk production of experimental multiplicities in light and heavy ion reactions at the LHC energies 
$\sqrt{s_{NN}} \sim 5$~TeV can be achieved, down to centralities $c \lesssim 1\%$, in an independent particle production framework, 
provided the following basic features are incorporated:

(i) Partonic degrees of freedom in the Glauber~\cite{Bialas:1977en,Bialas:1980zw,Bialas:2006kw} modeling of the initial state, with a few partons per nucleon (four or five partons both work in our study). 
In the past such a framework was successfully used at lower 
energies~\cite{Nouicer:2006pr,Bozek:2016kpf,Barej:2017kcw,Tannenbaum:2017ixt}. In partonic models, the deposition of entropy in the initial fireball increases faster 
with multiplicity compared to the models with the nucleon degrees of freedom.

(ii) Fluctuations in the entropy deposition by each wounded parton, generally induced by 
the quantum nature of the collision.
Overlaid fluctuations in the Glauber approach 
were considered before~\cite{Broniowski:2007nz,ALICE:2014xsp,Welsh:2016siu,Loizides:2016djv,Bozek:2017elk,ALICE:2018wma,ALICE:2025woy} as a necessary ingredient, 
in particular in the most central collisions or in {\it small} systems, e.g., $p$-Pb~\cite{Bozek:2013uha}. 

By {\it universality} of the description we mean that each wounded parton deposits the entropy in the same way, i.e., with the same distribution, 
irrespective of the collision system or centrality (the physical model parameters have the same values for all the studied reactions). 
The parameters of the distribution may, and do depend on the collision energy.\footnote{In the present case the differences in $\sqrt{s_{NN}}$ are negligible in that regard.}

Wounded partons deposit entropy in the initial fireball, and in turn this initial entropy is proportional to the final  number of the produced particles. In particular, this assumption is well satisfied in  common hydrodynamic models of heavy-ion collisions, involving  entropy production in the viscous hydrodynamic phase, hadronization, hadron rescattering and decays.  Note that our discussion does not use a particular model of the collision dynamics.
The universality argument is grounded in the basic assumption that the same number of initial sources should result in approximately the same number of particles, at a given energy, irrespective of its creation mechanism. This essentially means that the probability of producing a certain multiplicity is proportional (with same multiplicative factor) to  the initial  source strength (entropy) across all systems and centralities. A consideration of the full multiplicity distributions (and not only the mean multiplicity in several centrality bins)  for the considered collision systems (O+O, Ne+Ne, Xe+Xe, and Pb+Pb) is essential to constrain the parameters of the models and to assess its limits of applicability.

The relation between the initial  source strength and the experimental centrality definition is crucial for the determination of the centrality bins and a practical comparison to model predictions. However, a difficulty resides in a precise definition of the hadronic cross sections, as used in models, and the experimental data for ultra peripheral collisions, which  can have large admixtures of other types of interactions, in particular, the electromagnetic effects~\cite{ATLAS:2026zgq}. For model calculations involving the strong interactions only, it may result in a system dependent normalization of the overall cross section~\cite{Nijs:2021clz}.
To remedy the difficulty, it is customary to choose a certain range of centrality covering a majority of the particle production which can be assumed to be governed by purely strong interactions.
In our analysis we consider centralities 1-80\% for all the four studied systems.

\bigskip

{\bf Wounded Parton Model.}
Modeling the initial entropy deposition in the fireball (in the transverse plane) has a long history. 
Approaches based on the Glauber model~\cite{Glauber:1955qq,Miller:2007ri} played a major role, such as the Wounded 
Nucleon Model (WNM)~\cite{Czyz:1969jg,Bialas:1976ed}  (wounded
means a participant that collided {\it inelastically} at least once), a mixed model, where the wounded nucleons are amended with a fraction of binary collisions~\cite{Kharzeev:2000ph,Kharzeev:2001gp,Loizides:2016djv}, or extensions with quark/partonic degrees of freedom~\cite{Bialas:1977en,Bozek:2016kpf,Bialas:1980zw,Moreland:2018gsh,Nijs:2020roc}. 

In the Wounded Parton Model (WPM), a nucleon consists of $N_p$ partons, which can collide inelastically and deposit entropy in the initial state. 
In GLISSANDO 3~\cite{Bozek:2019wyr}, used in this work, the partons are distributed inside the nucleon with an exponential density,
\begin{eqnarray}
    \rho(r) \propto r^2 \exp \left[ -\sqrt{\frac{2}{3} \left( 1-\frac{1}{N_p}\right)} \frac{r}{r_0(s;N_p)} \right],
    \label{eq:nucldist}
\end{eqnarray}
where $r_0(s;p)$ controls the size of the nucleon built of partons, $s$ is the collision center-of-mass energy. In the wounded parton picture, the parton-parton inelastic collision profile has a Gaussian shape, 
\begin{eqnarray}
    p_{in}^{qq}(s,b) = \exp \left[- \frac{\pi b^2}{\sigma_{in}^{qq}(s;N_p)}\right],
    \label{eq:inelprof}
\end{eqnarray}
where $b$ is the impact parameter and $\sigma_{in}^{qq}$ is the parton-parton cross section. The parameters $r_0$ and $\sigma_{in}^{qq}$ are chosen in such a way that the COMPETE parametrization of the experimental $pp$ data~\cite{ParticleDataGroup:2016lqr} are accurately reproduced in GLISSANDO simulations. The nucleons in the nuclei are distributed according to the Woods-Saxon (WS) distribution for Pb+Pb and Ne+Ne, the Harmonic Oscillator distribution for O+O, and a deformed WS for Xe+Xe.

At lower collision energies, it has been found that the wounded quarks~\cite{Bialas:1977en}, i.e., three partons per colliding nucleon, 
lead to approximate scaling~\cite{Bozek:2016kpf,Barej:2017kcw} for various collision systems, where {\it scaling} means 
that the number of the produced particles is proportional to the number of the wounded objects with a proportionality constant depending only on the collision energy.  More partons (4--5) are needed to scale the Pb-Pb and $p$-Pb pseudo-rapidity spectra at $5.02$~TeV~\cite{Rohrmoser:2019xis}. 

Each wounded parton may deposit a varying entropy, due to quantum fluctuations. Additional fluctuations of the final multiplicity may occur in the dynamical evolution, e.g., during hadronization or resonance decays.  These effects of varying entropy deposition are implemented in GLISSANDO, or in other state-of-the-art Glauber-like models, e.g., TRENTO~\cite{Moreland:2014oya}. Typically, in studies of the multiplicity spectra, one uses the negative binomial distribution (NB)~\cite{ALICE:2013hur,ALICE:2014xsp,Bozek:2016kpf}, which is 
overlaid over the distribution of the wounded objects. In particular, we take 
\begin{eqnarray}
S=\sum_i   P_{\rm NB}(n | 1, \kappa), \label{eq:Sdef}
\end{eqnarray}
where $i$ runs over the wounded partons, and 
\begin{eqnarray}
    P_{\rm NB}(n | \bar n, \kappa) = \frac{\Gamma(n+\kappa)}{\Gamma(n+1) \ \Gamma(\kappa)} \frac{\kappa^\kappa \bar n ^ n}{(\bar n + \kappa)^{n+\kappa}} \label{eq:kappa}
\end{eqnarray}
is the NB distribution with the mean $\bar n=1$ and 
variance $\frac{\bar n ^2}{\kappa}+\bar n$.
The variable $S$ is the relative deposited strength of the entropy, 
proportional to the multiplicity of an event~\cite{Bozek:2019wyr,Moreland:2014oya}. 
By construction $S$ is a discrete variable which can assume value $0$. Such events cannot contribute to particle production, hence are discarded from the event samples.

\bigskip

{\bf Methodology of fitting the data.}
\begin{figure*}[!htbp]
\includegraphics[width=0.495\textwidth]{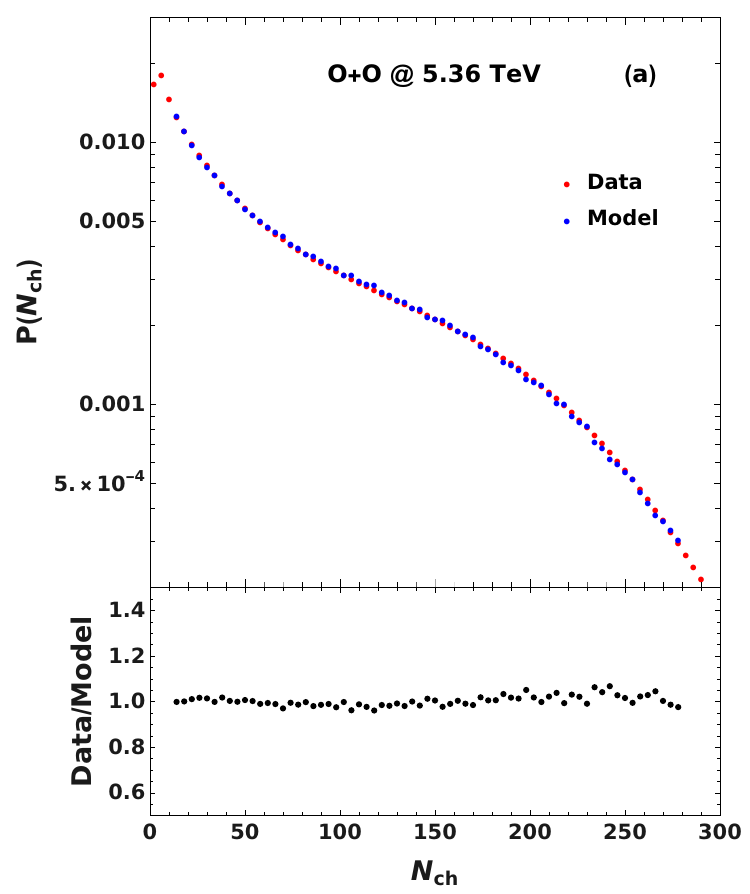}
\includegraphics[width=0.495\textwidth]{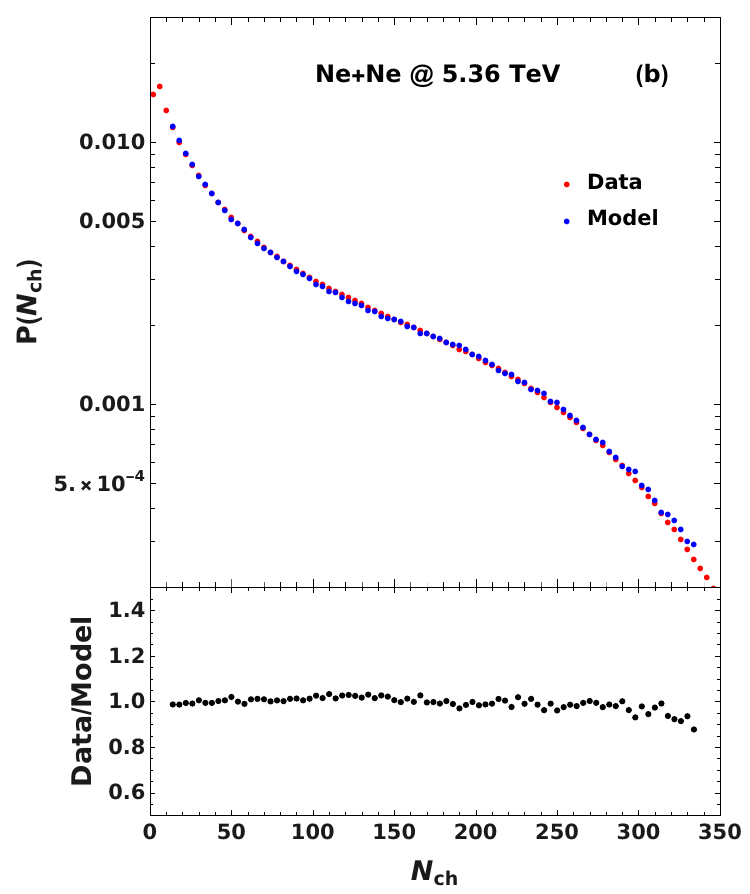} \\
\includegraphics[width=0.495\textwidth]{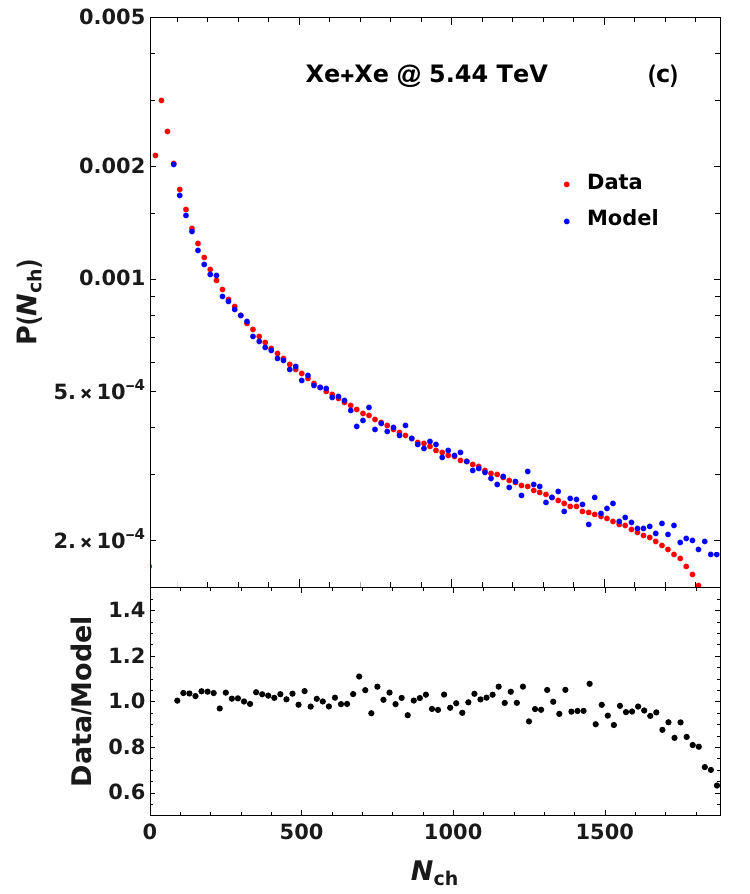}
\includegraphics[width=0.495\textwidth]{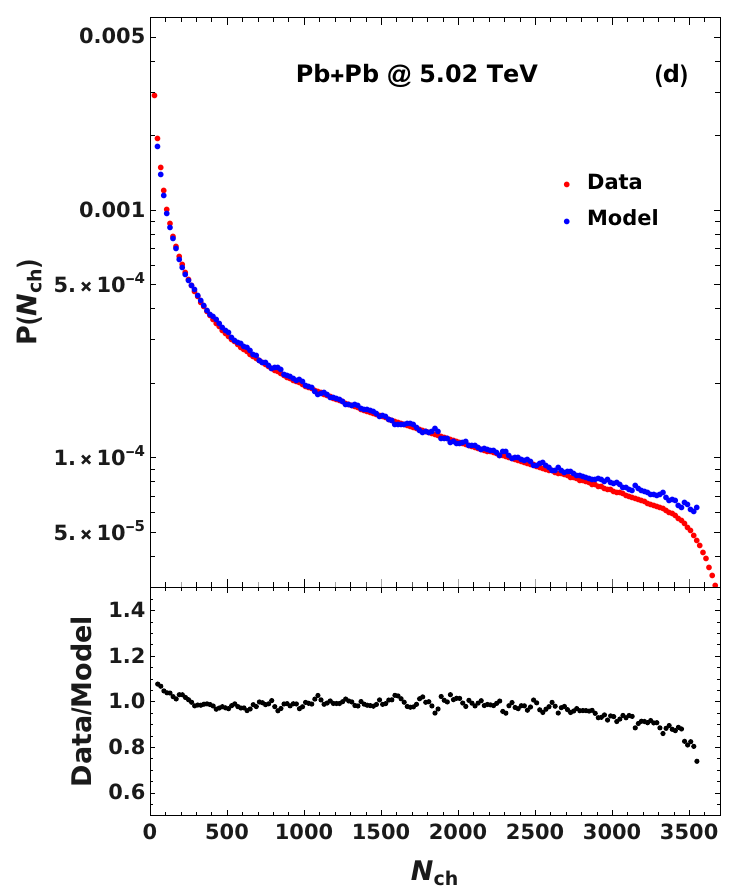} 
\caption{Multiplicity distributions as observed in experiments (red) and obtained by performing the joint $\chi^2$ fit of the WPM+NB to the data~\cite{ATLAS:2025nnt,ATLAS:2022dov} (blue) for O+O at $\sqrt{s_{NN}}=5.36$ TeV(a), Ne+Ne at $\sqrt{s_{NN}}=5.36$ TeV(b), Xe+Xe at $\sqrt{s_{NN}}=5.44$ TeV (c) and Pb+Pb at $\sqrt{s_{NN}}=5.02$ TeV (d) collision energies at LHC. The bottom part of each panel represents the data to model ratio denoted by black points.~\label{fig:dist_fit}}
\end{figure*}
Experimentally, for each reaction a multiplicity distribution (histogram) is recorded, denoted as $P^{\rm data}(N_{\rm ch})$.
At very low multiplicities, i.e., in the most peripheral 
collisions, there is a known issue associated with the Coulomb amplitude, becoming singular at $b\to\infty$ and interfering with the strong-interaction processes, which are
of interest here. A discussion how the problem is tackled can be found in~\cite{ALICE:2013hur}. There are also 
detector effects at the boundaries of acceptance, as well as effects related to efficiency. For these reasons, in routine model comparisons the suitably corrected 
experimental data are being used.

However, passing from the recorded to the corrected data in a model requires making a choice of the point where the model centrality is matched to the experimental distribution. 
Moreover, the corrected multiplicity 
histogram is being normalized, 
and then the corresponding centrality classes (quantiles) are determined from this distribution. 
Any uncertainty in modeling the Coulomb or the detector effects, in particular at the lowest $N_{\rm ch}$ end, subsequently carries over to the normalization and to 
the boundaries of the determined centrality bins. 
The difficulty can be  seen, e.g.,  in the recorded data histograms (red points) in Fig.~\ref{fig:dist_fit}, as the uncertainty in identifying the points close to  $N_{\rm ch}\to 0$ to the physical mechanisms included in the model  (note the log scale on the $y$ axis, 
visually diminishing the large size of the effect).

\begin{figure}[tb]
\includegraphics[width=0.47\textwidth]{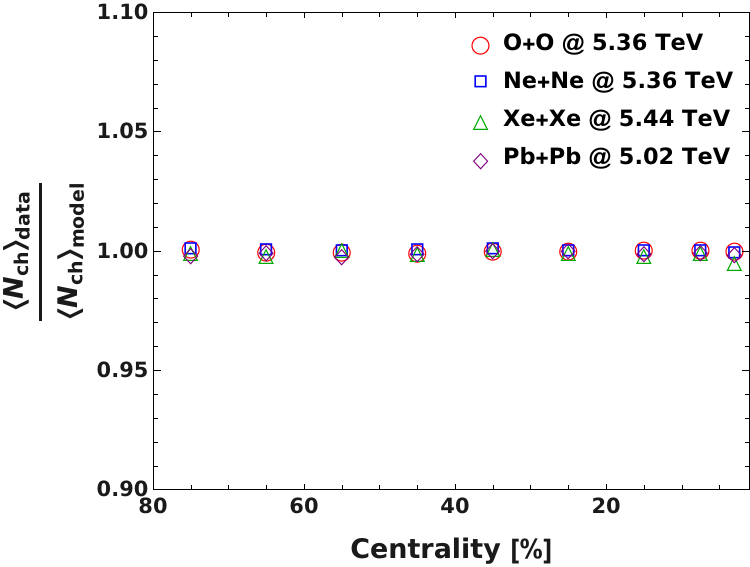} 
\caption{Data to model ratio of the mean multiplicity in centrality bins defined by the experimental multiplicity distribution. The red circle, blue square, green triangle and purple diamonds represent O+O, Ne+Ne, Xe+Xe and Pb+Pb collisions respectively at LHC energies.\label{fig:ratio}}
\end{figure}

To circumvent the difficulties discussed above, in our study we do not use the centrality classes 
as provided by the experimental analyses, but use directly the recorded $P^{\rm data}(N_{\rm ch})$ distributions, and decide to exclude the
most peripheral (lowest $N_{ch}$) range of the data from the fit to the model. In the present work we use uncorrected ($N_{\rm ch}^{\rm recorded}$) data from the 
ATLAS Collaboration~\cite{ATLAS:2022dov}.

We use the GLISSANDO~\cite{Bozek:2019wyr} implementation of WPM+NB model to generate $10^6$ events for each reaction $A$+$A$=O+O, Ne+Ne, Xe+Xe, and Pb+Pb. 
For each case, the obtained 
distribution of the variable $S$ is denoted as $P_{AA}^{\rm mod}(S)$. 
Both experimental and model histograms are conventionally normalized to unity, 
\begin{eqnarray}
\sum_{N_{ch}} P_{AA}^{\rm data}(N_{\rm ch})=1, \;\;\; \sum_S P_{AA}^{\rm mod}(S)=1.
\end{eqnarray}
The multiplicity is proportional to the deposited initial entropy, with 
the relation $N_{\rm ch}=\beta S$.
Then the 
model distribution transforms as $dS\, P_{AA}^{\rm mod}(S) = dN_{\rm ch} /\beta P_{AA}^{\rm mod}(N_{\rm ch}/\beta)$. To take into account the 
uncertainty of the normalization of the experimental histogram due to the above discussed effects, we introduce for each reaction a 
multiplicative parameter $\gamma_{AA}$ and compare both distributions in a chosen range of $N_{\rm ch}$ in the 
$\chi^2$ sense,
\begin{eqnarray}
    P^{\rm data}_{AA}(N_{\rm ch}) \simeq \frac{\gamma_{AA}}{\beta} P_{AA}^{\rm mod}\left (\frac{N_{\rm ch}}{\beta}\right).
\end{eqnarray}
Specifically, we construct a $\chi^2$-like quantity
\begin{eqnarray}
\hspace{-3mm}     \overline{\chi}^2_{AA}(\beta,\gamma,\kappa) = 
    \sum_i \frac{\left[ P^{\rm data}_{AA}(N_i) - \frac{\gamma_{AA}}{\beta} P_{AA}^{\rm mod}\left (\frac{N_i}{\beta}\right) \right]^2}{\sigma^2_{i, data}+\sigma^2_{i, mod}},
    \label{eq:chi2}
\end{eqnarray}
where, with our choice, for each reaction $i$ runs over all the experimental values of $N_i$ between centrality 80\% and 1\%.

The weights in Eq.~(\ref{eq:chi2}) involve both the experimental error, which is Poissonian, $\sigma^2_{i, data}=N_i$, and the model simulation error, 
which is also Poissonian and actually larger from the experimental one due to a smaller data sample, $\sigma^2_{i, mod}=N_i/\beta$. 

To perform a joint global fit we define the following function 
\begin{eqnarray}
   \overline{\chi}^2(\beta,\{\gamma_{AA}\},\kappa) = \sum_{AA} \overline{\chi}^2_{AA}(\beta,\gamma_{AA},\kappa) 
    \label{eq:chi2tot}
\end{eqnarray}
where the sum runs over the four collision systems: O+O, Ne+Ne, Xe+Xe and Pb+Pb, $\{\gamma_{AA}\}$ represents four normalization factors corresponding to the four reactions, resulting in six parameters in total. The expression of Eq.~(\ref{eq:chi2tot}) is minimized 
over all the six variables to obtain the best fit.\footnote{Due to the presence of the model errors, Eq.~(\ref{eq:chi2tot}) is not the $\chi^2$ distributions, hence standard
formulas for hypothesis testing of determining the model errors do not hold.}  

\bigskip

{\bf Results.}
The result of the fit are best for the variant with four partons per nucleon, $N_p=4$.\footnote{We have checked that  $N_p=3$ results in a significantly 
larger $ \overline{\chi}^2/{\rm DOF}$, whereas $N_p=5$ gives only slightly larger $\overline{\chi}^2/{\rm DOF}$ consistent with 1 and essentially equivalent description as 
for $N_p=4$.}
The optimum parameters are
\begin{eqnarray}
&& \gamma_{\rm OO}=0.999, \;\; \gamma_{\rm NeNe}=1.002, \nonumber \\ && \gamma_{\rm XeXe} =1.27, \; \; \gamma_{\rm PbPb}=0.86. 
\end{eqnarray}
Note large normalization correction for Xe+Xe and Pb+Pb case, while for O+O and Ne+Ne they happen to be properly normalized.
The best-fit values of other two parameters in Eq.~(\ref{eq:chi2tot}) are
\begin{eqnarray}
\beta=2.51, \;\; \kappa=0.23, 
\end{eqnarray}
with the value of $\overline{\chi}^2$ at the minimum giving $\overline{\chi}^2/{\rm DOF} = 0.86$. 

In Fig.~\ref{fig:dist_fit} we show our best fit model comparison to the data 
from ATLAS Collaboration for OO, NeNe~\cite{ATLAS:2025nnt} in panels (a) and (b), and for PbPb and XeXe~\cite{ATLAS:2022dov} in panels (c) and (d). 
In each panel we also show the data to model ratio at the bottom, in the linear scale. 
From panels (a) and (b) it can be seen that the model reproduces to within 5\% or better the data 
for O+O and Ne+Ne within the used centrality range 1-80\%.  For Xe+Xe and Pb+Pb reactions, as seen from panel (c) and (d), the model reasonably describes 
the data at lower multiplicities, but systematically overshoots the data, at large multiplicities, i.e.,  below $\approx$ 3\% centrality. 
This may be caused by our use of the uncorrected data, where the efficiency correction might be larger at higher multiplicities, of an unaccounted for 
physics effects, e.g., saturation.

In Fig.~\ref{fig:ratio} we show the model-to-experiment comparison in a way typically used in other studies, where average multiplicities are evaluated in centrality bins and  the ratio taken. Here we use the $N_{\rm ch}$ boundaries as obtained by us from the experimental distributions for centralities 1-5\%, 5-10\%, 10-20\%, \dots, 70-80\%. We note that the ratio is consistent with unity for all centrality windows and all collision systems, indicating the wounded parton scaling.

\bigskip

{\bf Conclusions.} 
Our results show that the wounded-parton model with 4 or 5 wounded partons per nucleon properly 
describes the average particle production across collisions with different system sizes at $\sqrt{s_{\rm NN}}\simeq 5$~TeV, with some discrepancy appearing for the most central 
heavy-ion collisions.
Thus, modeling of the initial entropy deposition  within the wounded parton framework can be used as a basis for the initial conditions in hydrodynamic calculations across very different systems, such as the light-ion and heavy-ion collisions at the LHC. 
The discrepancy for the most central collisions hints at the possibility of saturation-like effects in the most dense collision systems~\cite{Schenke:2012wb}.

\bigskip

{ \bf Acknowledgements.} RS thanks Somadutta Bhatta for discussion on the ATLAS data and Tribhuban Parida for valuable physics discussion. The authors acknowledge support from the Polish National Science Center grant 2023/51/B/ST2/01625.

\bibliography{ref}

\end{document}